# Consistency analysis and nuclear data validation for two series of beryllium reflector critical benchmark experiments[*]

Shengli CHEN, Tianxiang WANG

Sino-French Institute of Nuclear Engineering and Technology, Sun Yat-sen University, Zhuhai 519082, China

**Abstract**

Beryllium metal and beryllium oxide are important nuclear materials, with neutron-induced nuclear reaction data on beryllium playing a crucial role in nuclear energy research and development. Macroscopic validation is an essential step in the nuclear data evaluation process, providing a means to assess the reliability and accuracy of such data. Critical benchmark experiments serve as the most important references for this validation. However, discrepancies have been observed in two closely related series of beryllium-reflector fast-spectrum critical benchmark experiments, HMF-058 and HMF-066, which are widely used in current nuclear data validation. A previous systematic study indicates that these two series of experiments reach contradictory conclusions in validating the neutron-induced nuclear reaction data of beryllium, creating ambiguity in improving beryllium nuclear data. As a result, the total of 14 experiments in these two series cannot currently support high-accuracy validation of nuclear data. Although most researches on nuclear data validation and adjustment mainly focus on cross sections, the angular distribution of emitted neutrons is a key factor in reactor physics calculations. In this work, we address these inconsistencies by improving the secondary angular distributions of the (*n*, *n*) and (*n*, 2*n*) reactions of beryllium, thereby making the theoretical calculations (*C*) and experimental results (*E*) of these two series more consistent, and reducing the cumulative $\chi^2$ value from 7.58 using the ENDF/B-VII.1 evaluation to 4.52. All calculations based on the improved nuclear data agree with the experimental measurements within $1\sigma$ experimental uncertainty. With these enhancements, the consistency between the HMF-058 and HMF-066 series cannot be rejected within the $1\sigma$ experimental uncertainty. Based on the latest comprehensive evaluation of uranium nuclear data, this consistency is slightly improved, and the cumulative $\chi^2$ value decreases to 4.36 once again. Despite these advances, systematic differences in the expected values of *C*/*E* between the two series still exist. The *C*/*E* values of the HMF-066 series are generally 230–330 pcm lower than those of the HMF-058 series,

---

[This is an iFLYTEK AI translation, preliminarily proofread by the authors]

comparable to their experimental uncertainties of 200–400 pcm. Therefore, drawing a definitive conclusion about this systematic difference remains challenging. If the current improvement of differential nuclear data based on experimental data of $^9$Be is accurate, then the HMF-058 series experiments seem to be more reliable than the HMF-066 series. Ultimately, to achieve this goal, either reducing experimental uncertainty or designing and executing higher-precision integral experiments is required.

# 1. Introduction

Beryllium (Be) is the lightest non-gas element except lithium (Li). It has good physical, chemical, and nuclear reaction properties. For the slowing down of fast neutrons, the efficiency increases with decreasing atomic weight of the moderator. Because Li is too chemically active, Be is one of the best moderating materials among the available solid moderators. $^9$Be is the only naturally occurring isotope of Be. $^9$Be has a large thermal neutron scattering cross section and a small ratio of thermal neutron absorption to scattering, so it can be used as a good neutron reflector. In addition, $^9$Be can be used as a neutron multiplier since it can realize neutron multiplication through the (*n*, 2*n*) nuclear reaction. Therefore, metal Be has become a key nuclear material of great concern. In addition to metal Be, its oxide form BeO is also widely used in nuclear reactors [1], especially because of its high thermal conductivity, which allows for being doped in nuclear fuel to improve the thermal conductivity [2,3] of the fuel. Therefore, the study of neutron reaction characteristics related to Be is quite important.

Due to the lack of a systematic microscopic mechanism model to describe neutron-induced nuclear reactions in the full energy region, the current applied research concerning nuclear reactions is mainly based on nuclear reaction data. Given the large number of nuclides, many reaction channels, many physical quantities related to nuclear reactions (including reaction cross sections, angular distributions, energy distributions, etc.), and the dependence on incident energy, the commonly used nuclear reaction data are evaluated, validated, and then collected in nuclear databases. There are five general-purpose nuclear data libraries in the world, namely the American ENDF/B [4], European JEFF [5], JENDL [6] of Japan, CENDL [7,8] of China, and ROSFOND [9] of Russia, among which the ENDF/B evaluations are the most widely used in the world.

Macroscopic validation is a key step to ensure the reliability of evaluated nuclear reaction data. It employs the microscopic neutron reaction data to calculate macroscopic physical quantities, and the accuracy of microscopic nuclear data is evaluated by comparing the calculation results with the experimental results. The target physical quantities of macroscopic validation, also known as integral quantities, are usually neutron spectrum and reactor physical parameters, of which the most typical is the experimental value of effective multiplication factor ($k_\text{eff}$) collected in the International Criticality Safety Benchmark Evaluation Project (ICSBEP) [10]. The comparison between the calculated ($C$) and experimental values ($E$) of $k_\text{eff}$ has become a standard procedure for the validation of evaluated nuclear data. Based on ICSBEP, China has also established a computational model library, ENDITS-2.1 [11], containing 2237 critical benchmark experiments to further promote the verification of nuclear data.

However, a previous study has found that the highly similar HMF-058 and HMF-066 series of beryllium reflector critical benchmark experiments give inconsistent conclusions in nuclear data validation, so there may be systematic problems in the measurement or evaluation of at least one of the series of benchmark experiments [12]. In addition, the calculated results of the HMF-058 series based on the latest evaluated nuclear data at that time are higher than the experimental results by more than $1\sigma$ experimental uncertainty [12]. In view of the wide application of these two series of critical benchmark experiments in the macroscopic validation of nuclear data libraries, systematic analysis and evaluation of them are essential for nuclear data evaluation and subsequent applications. The present study greatly improves the agreement between HMF-058 and HMF-066 by improving key nuclear data, especially those of the neutron scattering angular distribution, which has been paid less attention to before. The final results of both series are within $1\sigma$ experimental uncertainty.

## 2. Critical benchmark experiment

2.1 Experimental setup and parameters

As of 2022, the ICSBEP Handbook has included 598 series of 5,159 critical benchmark experiments [13]. Each experiment in each series in the manual has a corresponding number, including a three-tier letter abbreviation (nuclear fission fuel type-fuel physical form-neutron spectrum) indicating the characteristics of the experiment, followed by a series of numerical numbers, and an experiment numerical number within each series. This study focuses on two series of beryllium reflector critical benchmark experiments, with ICSBEP manual numbers HEU-MET-FAST-058 and HEU-MET-FAST-066, where HEU stands for highly enriched uranium, MET stands for metal fuel, FAST stands for fast neutron spectrum, and 058 and 066 are specific numbers under the HEU-MET-FAST category. For critical benchmark experiments in each series, ICSBEP appends a sequential number starting from 001 to the series name, such as HEU-MET-FAST-058-001. For simplification, HMF-058 and HMF-066 are commonly used to refer to these two experimental series.

Five experiments of the HMF-058 series were carried out between 1958 and 1959, while nine experiments of the HMF-066 series were carried out in 1960. The two series of experiments are designed in spherical geometry. The HMF-058 series is composed of a beryllium sphere (radius 4.98 mm), nickel shell (outer diameter 5.21 mm), void layer (outer diameter 5.56 mm), fuel shell, and beryllium reflector shell from inside to outside. Different experiments only change the thickness of the fuel shell and beryllium reflector shell to make the device reach critical state; The HMF-066 series is highly similar to the HM-F058, with only one beryllium shell added between the void and the fuel layer, and the thickness of the inner and outer beryllium reflectors and the fuel layer changed for different experiments. Fig. 1 shows the schematic of HMF-058-002 (radius 152.11 mm) and HMF-066-007 (radius 151.49 mm) with the closest overall dimensions. The simple geometric and material designs of these two series also make them widely used in macroscopic validation of nuclear data.

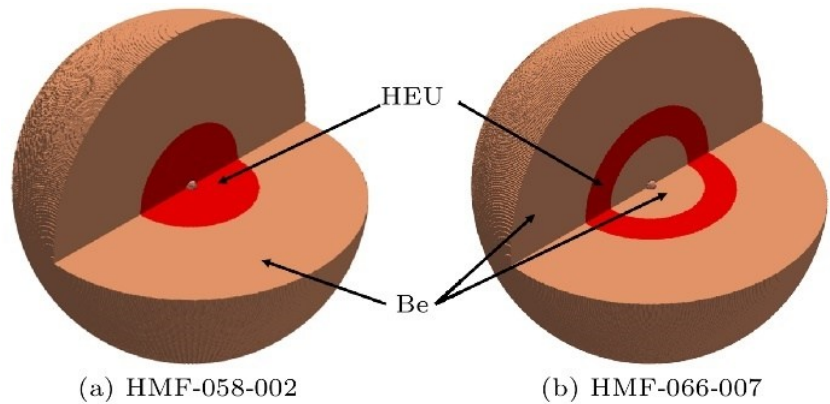

(a) HMF-058-002    (b) HMF-066-007

**Figure 1.** Schematic of HMF-058-002 and HMF-066-007 experiments (the sphere in the core represents the Be core - Ni layer that remains unchanged in different experiments).

2.2 Experimental results

With reference to the previous study [12], the experimental $k_{eff}$ values are shown in Fig. 2 according to the thicknesses of the beryllium reflectors. Since the HMF-066 series has two layers of beryllium reflectors, the thickness of the outer reflector (consistent with the HMF-058 series) and the total thickness of the inner and outer reflectors are shown as variables. It is worth noting that the five benchmark experiments of the HMF-058 series are all critical, while the benchmark experiments of the HMF-066 series are slightly supercritical. The theoretical calculations of $k_{eff}$ based on the evaluated nuclear data involved in this study are carried out by the OpenMC [14] Monte Carlo neutron transport code.

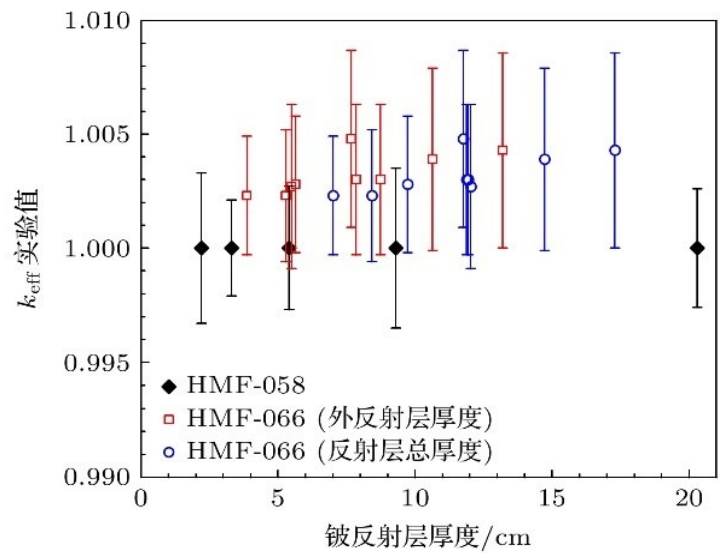

**Figure 2.** Experimental measurements and uncertainties of $k_{eff}$ for HMF-058 and HMF-066 series experiments versus the thickness of the Be reflector.

### 3. Improvement of calculation results

3.1 Improvement of HMF-058 results

The previous study has shown that the sensitivity coefficients of HMF-058 and HMF-066 for the $^9$Be reaction cross section are similar, so the agreement between them cannot be improved by modifying the reaction cross sections alone [12]. In fact, although the current nuclear data validation and adjustment studies mainly focus on the reaction cross section data, the angular distribution of the outgoing neutron is an important factor affecting the calculation of $k_{eff}$ [15]. For $^9$Be, as shown in Fig. 3, the secondary angular distributions (SADs) of the $(n, n)$ and $(n, 2n)$ reactions in the JENDL-4.0 evaluation [18] are generally better than those in ENDF/B-VII.1 [19], compared with the experimental results collected in EXFOR [16,17]. The accuracy of the calculation results based on better SAD data is higher. In addition, as shown in Fig. 4, the thermal neutron absorption cross section of $^9$Be in ENDF/B-VII.1 ( (10.03±0.50) mbarn) is significantly higher than the experimental values ( (8.4±0.1) mbarn on average), and its accuracy is not as good as the JEFF-3.3 evaluation ( (8.78±0.44) mbarn) [20]. Therefore, the $^9$Be neutron absorption cross section of ENDF/B-VII.1 is also replaced by that of JEFF-3.3, as in our previous work [20].

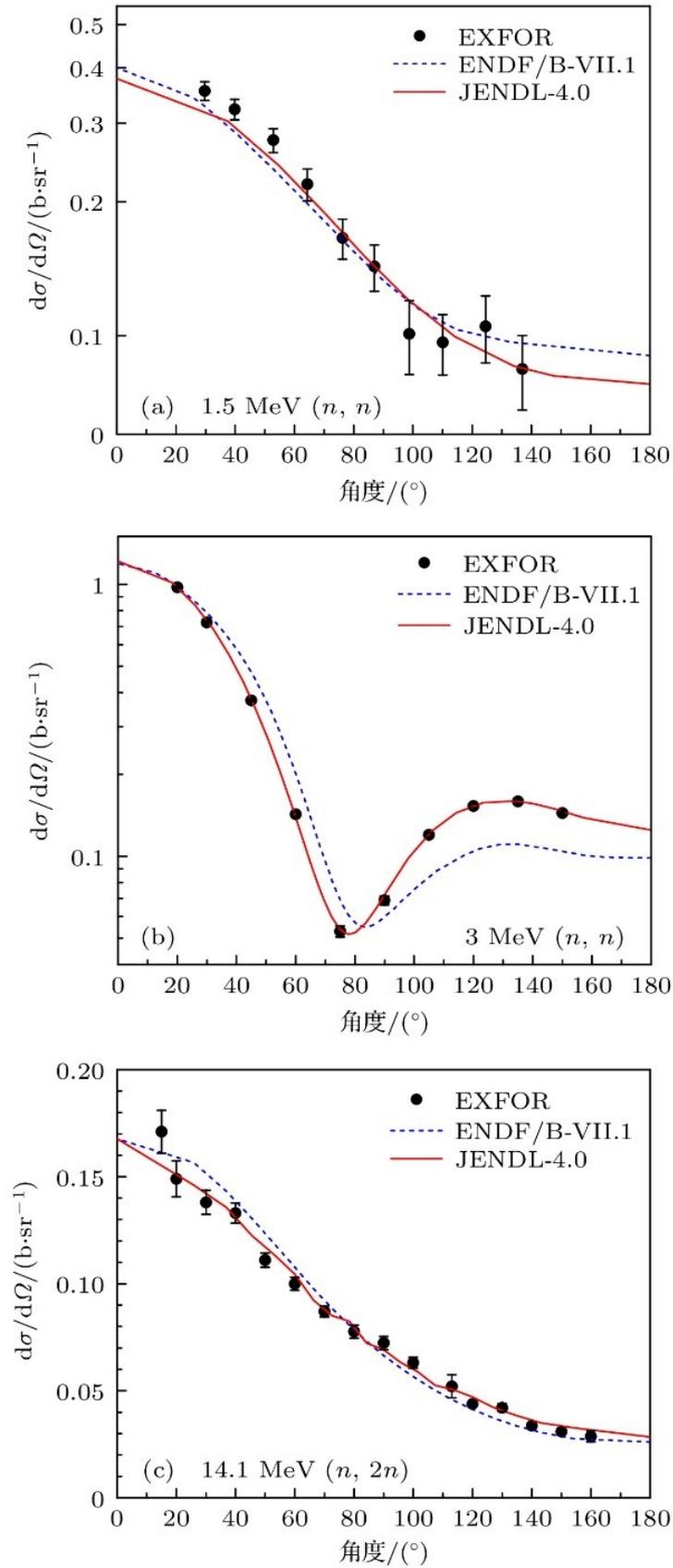

**Figure 3.** Angular distributions of outgoing neutrons for $^9$Be$(n, n)$ and $(n, 2n)$ reactions.

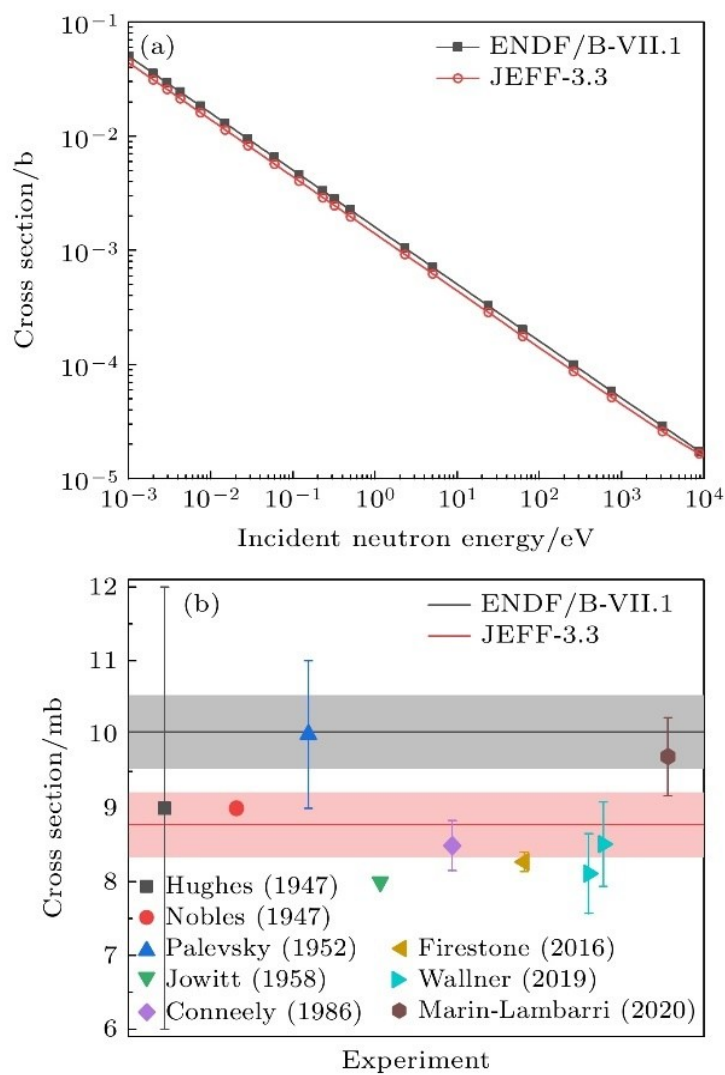

**Figure 4.** (a) Neutron capture cross section of $^9$Be; (b) the comparison with experimental data for thermal neutron [20].

Fig. 5 shows the ratio of calculated to experimental (*C*/*E*) results for the HMF-058 series. Based on the updated SAD data, the calculated values are significantly decreased by 70-180 pcm (average reduction of 120 pcm), and the calculated and measured values of the five experiments are consistent within 1$\sigma$ experimental uncertainty. The cumulative $\chi^2$ is reduced from 6.46 based on the ENDF/B-VII.1 evaluation to 2.52. In fact, the reduction of the neutron absorption cross section results in the increase in $k_{eff}$, the influence of SAD may be slightly larger than those shown in Fig. 5. It can be inferred that the overestimation of HMF-058 series benchmark experiments for ENDF/B-VII.1 is most likely caused by the SADs of (*n*,*n*) and (*n*, 2*n*) reaction channels of $^9$Be. The improvement of the $^9$Be angular distribution based on the differential experimental data is consistent with the improvement of the calculation accuracy of the integral quantity $k_{eff}$ of the critical benchmarks.

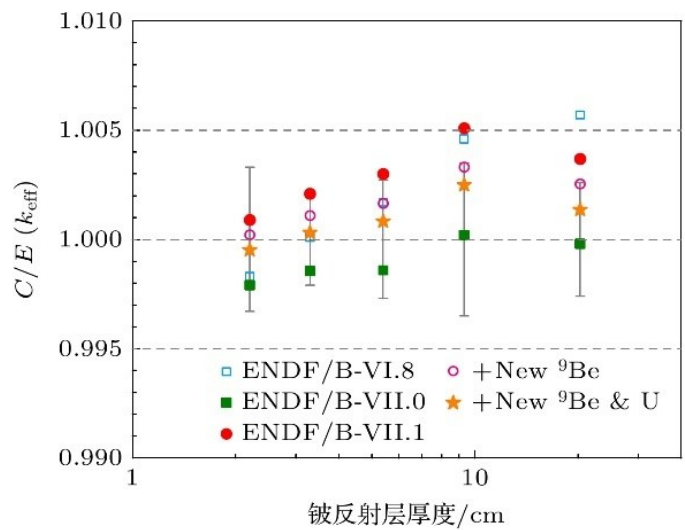

**Figure 5.** Ratio of calculation to experimental measurement (C/E ) of $k_{eff}$ for HMF-058 series: previous [12] and improved results.

Although the HMF-058 critical benchmark experiment is highly sensitive to $^9$Be nuclear data, its experimental setup with high enrichment $^{235}$U implies higher sensitivity to $^{235}$U nuclear data. After the release of ENDF/B-VII.1, the OECD/NEA coordinated the Collaborative International Evaluation Library Organization (CIELO) project to evaluate the nuclear data of hydrogen (H), oxygen (O), iron (Fe), uranium (U), and plutonium (Pu) isotopes, which have a great impact on the application of nuclear technology [21]. The complete set of nuclear data for uranium isotopes has been re-evaluated [22]. Although there is an international standard [23] for the neutron-induced fission cross section of $^{235}$U, other non-standard data still have a greater impact on the critical calculation, such as the fission neutron multiplicity measuring the average number of neutrons per fission. Therefore, the complete set of nuclear data for uranium (including $^{235}$U and $^{238}$U) evaluated by CIELO is used to calculate the HMF-058 series benchmarks. As shown in Fig. 5, the calculated results are reduced by 70-120 pcm (about 90 pcm on average), and the agreement between the calculated results and the experimental values is further improved. This conclusion is consistent with the reduction of fission neutron multiplicity in the newly evaluated $^{235}$U nuclear data [22], where the sensitivity of theoretical $k_{eff}$ to the latter is close to 1.

Because of its specific material design, the HMF-058 series benchmark experiments are only sensitive to the neutron-induced nuclear reaction data of $^9$Be and $^{235}$U. The improvement of the angular distribution of the emitted neutrons in $^9$Be and the re-evaluation of the complete set of $^{235}$U nuclear data have greatly improved the agreement between the calculated and measured $k_{eff}$, and the cumulative $\chi^2$ has been reduced from 6.46 based on the ENDF/B-VII.1 evaluation database to 0.92. This result shows that the differential experimental data, such as the relevant SADs, are consistent with the HMF-058 series integral critical benchmark experiments.

### 3.2 Consistency between HMF-058 and HMF-066

After updating the above nuclear reaction data, the calculated results of HMF-058 are in good agreement with the experimental measurements. However, the previous study has shown that there is a systematic deviation between HMF-058 and HMF-066, and improving the calculation results of HMF-058 will lead to an increase in the deviation between the calculation and experiment of HMF-066 [12]. The $C/E$ results of nine critical benchmark experiments of HMF-066 series based on updated nuclear data are presented in Fig. 6. Compared with the results based on the original ENDF/B-VII.1 data, the calculated results are systematically decreased, which is consistent with the conclusion of reference [12]. The decreases due to the updated $^9$Be and U data are 110-320 pcm (190 pcm on average) and 30-100 pcm (70 pcm on average), respectively.

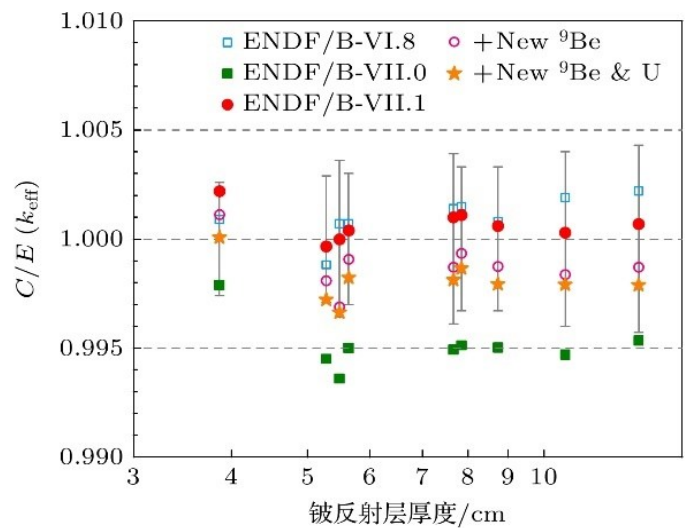

**Figure 6.** Ratio of calculation to experimental measurement ($C/E$) of $k_{\text{eff}}$ for HMF-066 series: previous [12] and improved results.

However, whether only updating the angular distribution and absorption cross section of $^9$Be or updating the complete set of nuclear data of U at the same time, the calculated and measured results of the HMF-066 series are basically in agreement within $1\sigma$ experimental uncertainty. The cumulative $\chi^2$ of 9 experiments increased from 1.12 based on the ENDF/B-VII.1 evaluation to 2.01 and 3.44, respectively. Therefore, in the view of statistics, it can still be considered that the calculation and measurement values of the HMF-066 series are consistent within $1\sigma$ experimental uncertainty.

In view of the high similarity between HMF-058 and HMF-066 series experimental designs, quantitative similarity indicators can be used for similarity analysis. In the previous study [12], the similarity (cosine value between two sensitivity coefficient vectors, also known as $E$ index [24] or cosine similarity) between five experiments of the HMF-058 series and nine experiments of the HMF-066 series has been systematically calculated. Taking HMF-058-001, HMF-058-002, and HMF-056-003 as examples, the present work selects the corresponding highly similar HMF-066-001 (99.4% similarity) and HMF-066-005 (99.3% similarity), HMF-066-006 (99.8% similarity) and HMF-066-002 (99.8% similarity), and HMF-066-003 (99.5% similarity), respectively [12], for comparison,

Fig. 7 shows the $C/E$ results of the $k_{eff}$ of the three sets of highly similar critical benchmark experiments. The previous results based on five different evaluated nuclear data libraries can be classified into two categories [12]: ENDF/B-VII.1, JENDL-4.0u, and JEFF-3.2 results are similar, and CENDL-3.1 and ROSFOND-2010 results are similar. The former overestimates the $k_{eff}$ of most HMF-058 series experiments, but is in good agreement with the $k_{eff}$ of HMF-066 series experiments; the latter is in good agreement with the $k_{eff}$ of HMF-058 series experiments, but underestimates the $k_{eff}$ of most HMF-066 series experiments. Based on the updated data of this study, although there are still differences between the two series, the consistency of all calculations and experimental results is guaranteed within $1\sigma$ experimental uncertainty.

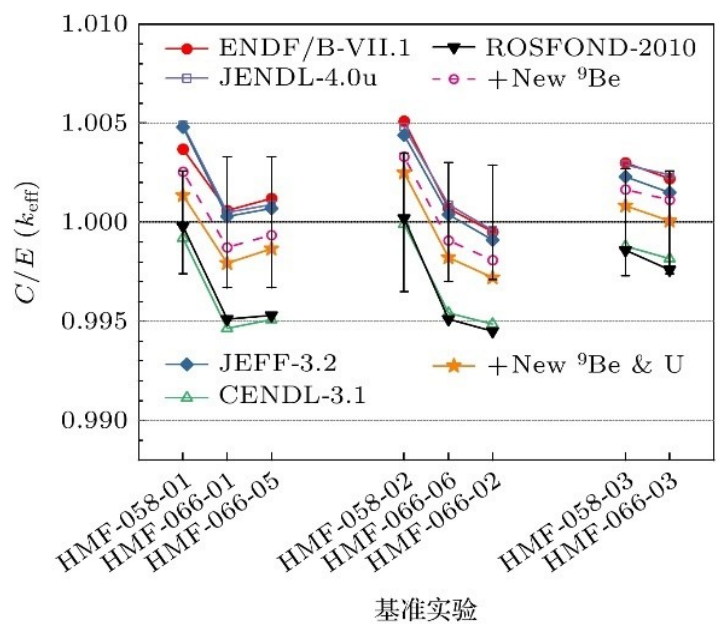

基准实验

**Figure 7.** Comparison of $C/E$ for $k_{eff}$ of highly similar experiments.

**4. Discussion**

Numerical simulations and calculations of critical benchmark experiments rely on a complete set of relevant nuclear reaction data. The sensitivity analysis of the calculation results on the reaction cross section has received extensive attention, including the sensitivity of some ICSBEP benchmark experiments included in the DICE database [25]. The angular distribution of emitted neutrons also has a great influence on the calculation of $k_{eff}$ [15], but there is little relevant research at present. One of the important reasons is that the cross section of each reaction varies only with the incident neutron energy, so the corresponding perturbation calculation and sensitivity expression are relatively easy. However, the angular distribution involves two free variables, the incident neutron energy and the outgoing neutron angle, so the complexity of the correlation analysis and representation is significantly increased. Since the effect of SADs on the $k_{eff}$ calculation is extremely sensitive to the reactor or critical benchmark experiment design, even though the qualitative prediction of this effect remains difficult for relatively complex systems [26]. Therefore, the influence of the angular distribution on the calculation of $k_{eff}$ cannot be ignored.

For the two series of beryllium reflector critical benchmark experiments of HMF-058 and HMF-066, the previous study based on cross-section sensitivity analysis has shown that there are systematic deviations between them [12]. In the present study, the accuracy of the original ENDF/B-VII.1 nuclear data for $^9$Be was improved by updating the angular distributions of the $^9$Be $(n,n)$ and $(n, 2n)$ reaction channels and neutron absorption cross sections. Based on the updated $^9$Be nuclear data, the calculated results of the two series are in good agreement with the experimental measurements within $1\sigma$ experimental uncertainty. The cumulative $\chi^2$ of 14 experiments in the two series reduces from 7.58 based on the ENDF/B-VII.1 evaluation to 4.52. Therefore, in a statistical sense, the consistency between the two series cannot be rejected within $1\sigma$ experimental uncertainty.

At that same time of update the $^9$Be nuclear data, replacing the full set of U (including $^{235}$U and $^{238}$U) nuclear data in ENDF/B-VII.1 by CIELO evaluations, the agreement between the calculated results of HMF-058 and the experimental measurement was further improved, and the agreement between the two series was also slightly improved (the cumulative $\chi^2$ was reduced from 4.52 to 4.36). However, the effect of U nuclear data improvement on HMF-058 and HMF-066 series is not as significant as that of $^9$Be nuclear data. Because the full set of CIELO nuclear data for U is almost all re-evaluated, including the fitting of the oscillation of fission neutron multiplicity as a function of the incident neutron energy [22], we did not quantify the impact of specific U nuclear data on the critical benchmark experimental calculations. However, this study still verifies that the complete CIELO set of U reaction nuclear data is generally better than ENDF/B-VII.1, so it is reasonable to be directly adopted by ENDF/B-VIII.0 database.

Of course, from the first two sets of results of Fig. 7, it can be seen that there are still systematic differences between the *C/E* expectations of HMF-058 and HMF-066 series, especially after updating the nuclear data, the *C/E* of HMF-066 is more deviated from 1. If the current nuclear data improvement based on the differential experimental data is correct, it can be inferred that the HMF-058 series experiments are relatively more reliable. However, because the calculated and measured results of the two series are consistent within 1$\sigma$ experimental uncertainty, and the cumulative $\chi^2$ of 14 experiments in the two series is only 4.36, it is impossible to reject that the two series are consistent within 1$\sigma$ experimental uncertainty in a statistical sense. If we want to get further conclusions about the consistency of the two series, we need to reduce the experimental uncertainty or carry out new experiments with higher precision.

## 5. Conclusions

The previous study based on several evaluated nuclear data libraries has shown that there are systematic deviations between the highly similar HMF-058 and HMF-066 series of beryllium reflector critical benchmark experiments [12]. The angular distribution of emitted neutrons has an important influence on the critical calculation [15], but there are only a few relevant studies. In the present study, by improving the angular distribution data of $^9$Be (*n,n*) and (*n*, 2*n*) reaction channels, it is found that the theoretical calculation results of the two series are consistent with the experimental measurement within 1$\sigma$ experimental uncertainty, so the agreement between them cannot be rejected within 1$\sigma$ experimental uncertainty. The agreement is further slightly improved by using the full set of U nuclear data evaluated by CIELO, which also illustrates the importance of nuclear data optimization for the consistency judgment of critical benchmark experiments.

At the same time, it should be noted that although the improved nuclear data make the calculated and experimental results of HMF-058 and HMF-066 series consistent within 1$\sigma$ experimental uncertainty, the overall expectation of *C/E* for HMF-066 is still 230-300 pcm lower than that of HMF-058. The improvement of the $^9$Be (*n,n*) and (*n*, 2*n*) reaction angular distributions and the complete set of U nuclear data based on the differential experimental data is consistent with the conclusion of macroscopic validation against the HMF-058 series integral experiments. The experimental uncertainties of the two series of critical benchmark experiments $k_{\text{eff}}$ are 200-400 pcm, which are comparable to the above differences and close to the uncertainty target propagated from the covariance of nuclear data in fast neutron reactor calculations [27]. Therefore, it is particularly important to reduce the experimental uncertainty or carry out more precise integration experiments for further analysis of the consistency of the two series of experiments and to test the accuracy of nuclear reactor calculations.

The authors acknowledge Dr. Zehua Hu (Institute of Applied Physics and Computational Mathematics) for the in-depth discussion, providing the data of reference [12], and revising this paper.

**References**

[1]   Hou M, Zhou X, Liu B 2022 Nucl. Eng. Technol. 54 4393

[2]   Chen S, Yuan C 2020 Nucl. Mater. Energy 22 100728

[3]   Chen S L, He X J, Yuan C X 2020 Nucl. Sci. Tech. 31 32

[4]   Brown D A, Chadwick M B, Capote R, et al. 2018 Nucl. Data Sheets 148 1

[5]   Plompen A J M, Cabellos O, De Saint Jean C, et al. 2020 Eur. Phys. J. A 56 181

[6]   Iwamoto O, Iwamoto N, Kunieda S, et al. 2023 J. Nucl. Sci. Technol. 60 1

[7]   Ge Z G, Chen Y G 2020 Nucl. Phys. Rev. 37 309

[8]   Ge Z, Xu R, Wu H, et al. 2020 EPJ Web Conf. 239 09001

[9]   Zabrodskaya S V, Ignatyuk A V, Koshcheev V N, et al. 2007 RUSFOND-Russian National Library of Evaluated Neutron Data https://vant.ippe.ru/en/year2007/neutron-constants/774-1.html

[10] Briggs J B, Scott L, Nouri A 2003 Nucl. Sci. Eng. 145 1

[11] Wu H C, Zhang H Y 2024 At. Energy Sci. Tech. 58 1271

[12] Hu Z H, Yin Y P, Ye T 2016 Acta Phys. Sin. 65 212801

[13] NEA 2024 "ICSBEP Handbook 2022-23", International Criticality Safety Benchmark Evaluation Project Handbook (Database) https://www.oecd-nea.org/jcms/pl_20291

[14] Romano P K, Horelik N E, Herman B R, et al. 2015 Ann. Nucl. Energy 82 90

[15] Chen S, Vandermeersch E, Tamagno P, et al. 2021 Ann. Nucl. Energy 163 108553

[16] Otuka N, Dupont E, Semkova V, et al. 2014 Nucl. Data Sheets 120 272

[17] Zerkin V V, Pritychenko B 2018 Nucl. Instrum. Methods Phys. Res. A 888 31

[18] Shibata K, Iwamoto O, Nakagawa T, et al. 2011 J. Nucl. Sci. Technol. 48 1

[19] Chadwick M B, Herman M, Obložinský P, et al. 2011 Nucl. Data Sheets 112 2887

[20] Wang T, Chen S, Xu S, et al. 2023 Ann. Nucl. Energy 192 110017

[21] Chadwick M B, Capote R, Trkov A, et al. 2018 Nucl. Data Sheets 148 189

[22] Capote R, Trkov A, Sin M, et al. 2018 Nucl. Data Sheets 148 254

[23] Carlson A D, Pronyaev V G, Capote R, et al. 2018 Nucl. Data Sheets 148 143

[24] Broadhead B L, Rearden B T, Hopper C M, et al. 2004 Nucl. Sci. Eng. 146 340

[25] Nouri A, Nagel P, Briggs J B, Ivanova T 2003 Nucl. Sci. Eng. 145 11

[26] Wang T X, Xu S Q, Li Z, Chen S L 2025 Ann. Nucl. Energy 210 110851

[27] Cabellos O, Hursin M, Palmiotti P 2023 EPJ Web Conf. 284 14012